\documentclass[twoside,reqno]{HERON}
\usepackage[utf8]{inputenc}
\usepackage{listings}
\usepackage{epsfig,cite,colordvi}
\usepackage{graphicx}
\usepackage{url}
\usepackage{times}
\usepackage[colorlinks=false]{hyperref}
\usepackage{makeidx}
\begin{document}

\title{Nuclear Dynamics with the Sky3D code}

\runningheads{Nuclear Dynamics with the Sky3D code}{P. D. Stevenson}

\begin{start}

\author{P. D. Stevenson}{1}

\index{P. D. Stevenson}

\address{Department of Physics, University of Surrey, Guildford, Surrey, GU2 7XH, UK}{1}

\begin{Abstract}
A description is presented of how to use the Sky3D time-dependent Hartree-Fock code to calculate giant monopole resonances.  This requires modification to the code, and a step-by-step guide of how to make the necessary modification is given.  An example of how to analyse the output of the code to obtain quantities of physics interest is included.  Together, the modifications and the post-processing are intended to serve as a typical example of how the code, which was designed to be extendable to particular users' needs, can be extended.
\end{Abstract}
\end{start}

\section{Introduction}
The time-dependent Hartree-Fock (TDHF) method \cite{Sim12} has, in the last few years, become a sufficiently mature technique that it may be used routinely in rather realistic situations describing a range of nuclear dynamics.  A recently-published code, \textit{Sky3D} \cite{Mar14} has made available a general purpose TDHF code, representing nuclei on a coordinate-space grid with no symmetry resetrictions.  The code uses the effective Skyrme interaction, making it in fact really a Time-Dependent Density Functional Code (TDDFT), though the name TDHF is so ubiquitous the literature even for such Skyrme-based approaches, that this slightly inaccurate nomenclature is kept.

The paper describing the code \cite{Mar14} is open access, and comes with a link to the \textit{Computer Physics Communications} program library whence the code can be downloaded, along with a substantial users manual, which goes into considerable detail about the program's inner workings.  That being the case, such details are not repeated here, but rather an example of out-of-the-box running (in the static case) is given, along with an example of extending the code, which will be required for some cases of interest.  The present paper should be read in conjunction with the paper describing the code.

Before embarking on the examples calculations, some mention will first be given about the kinds of application, with some references, that the code could be (and in some cases has been) applied to.  A more comprehensive review of nuclear TDHF and related methods furnishes further examples \cite{Sim12}.  Certain kinds of excited states of single nuclei may be explored.  The most common example is that of giant multipole resonance states in the linear regime \cite{Toh02,Ste04, Nak05, Uma05, Bri06, Ste07, Nak07, Ina09, Par13}.  More sophisticaed examples of giant resonance type calculations include studying nonlinearities and mixed strength functions \cite{Sim05,Alm05a}, skin vibrations \cite{God13}, spining toroidal configurations \cite{Ich12} and clustered excitations \cite{Ich11a,Ich11b}.  A considerable body of literature describes the use of TDHF for nuclear dynamics involving more than one body.  Some recent examples include cases of fusion \cite{Uma86,Uma14,Sim13}, transfer reactions \cite{Sim10,Sek13}, fission \cite{Uma10,God14,Sim14}, fragmentation \cite{Iwa10} and deep-inelastic collisions \cite{Ste12,Gol09}.  All such examples may be explored using the TDHF code, though in many cases, user-extensions to the code must be written.  An example of such an extension -- the case of giant monopole resonances -- is presented in this paper.

\section{Static calculations}
In order to perform a TDHF calculation, an initial state is required.  This is obtained by running the \textit{Sky3D} code in static mode, where it operates as a standard Skyrme-Hartree-Fock code, similar to other available codes such as \textit{HFODD} \cite{Sch12}, though much less fully-featured as such dedicated static solvers, as \textit{Sky3D} is geared only to providing simple starting states for time-dependent calculations. 

A suitable input file to generate the ground state of $^{18}$O is given below:

\small
\begin{verbatim}
 &files wffile='O18' /
 &force name='SkI4', pairing='NONE' /
 &main mprint=100,mplot=100,
  mrest=100,writeselect='r',
  imode=1,tfft=T /
 &grid nx=24,ny=24,nz=24,dx=1.0,dy=1.0,dz=1.0,
  periodic=F /
 &static nprot=8, nneut=10, 
  radinx=3.1,radiny=3.1,radinz=3.1,
  x0dmp=0.40,e0dmp=100.0,tdiag=T,tlarge=F,
  maxiter=2000,serr=2D-8 /
\end{verbatim}
\normalsize

The instruction {\tt imode=1} sets the operation of the code to be static Hartree-Fock.  A particular Skyrme force (SkI4 \cite{Rei95}) is chosen, and the number of protons and neutrons set.  For most purposes (except the inclusion of pairing), other parameters given in the input file for static runs can be left as in the file presented here.  

\begin{figure}[tb]
  \includegraphics[width=\textwidth]{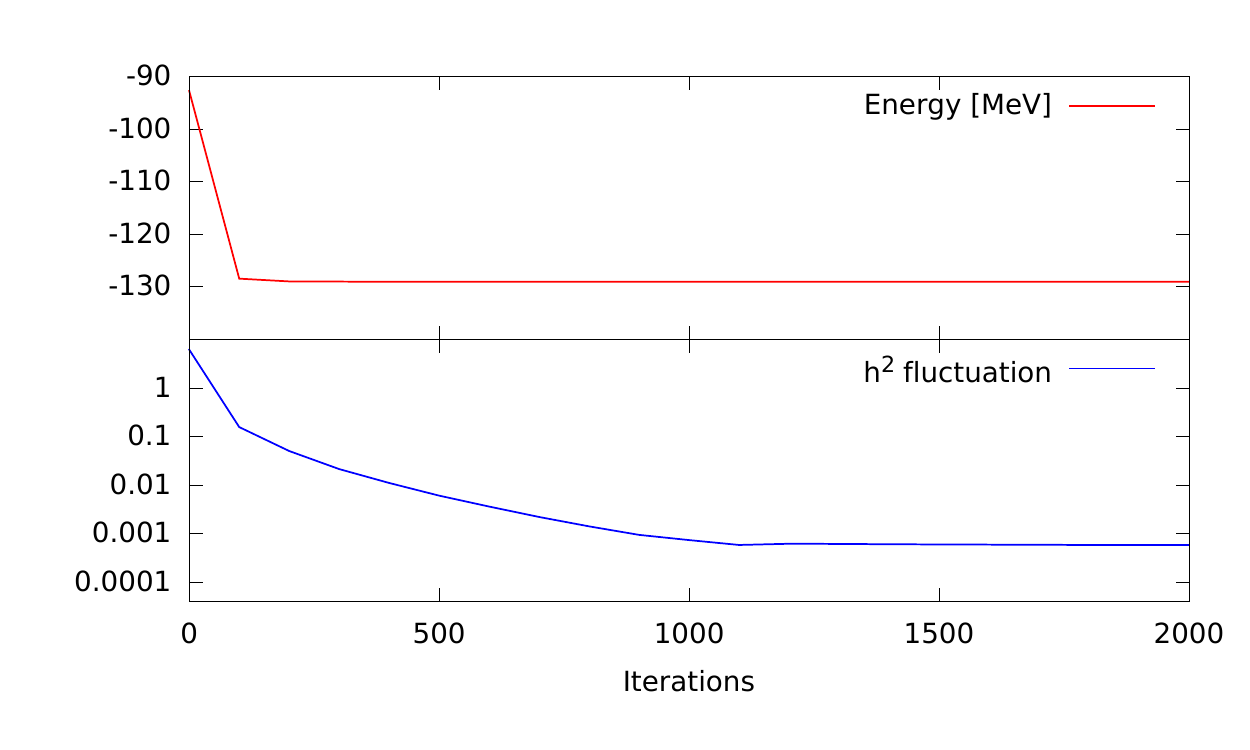}
  \caption{Convergence of the total Hartree-Fock energy (top panel) and in the fluctuations in the single particle wave functions (lower panel) for the case of $^{18}$O.\label{fig:conver}}
\end{figure}

The correct convergence of the wave functions to the true Hartree-Fock states in the static run can be checked in the file {\tt conver.res}.  One should take some care to ensure that not only has the total Hartree Fock energy converged, but that the fluctuations in the the single particle states are converged to small numbers.  Figure \ref{fig:conver} shows the convergence of the total energy for the input file given above, along with the average single particle energy fluctuation as given by the expression
\begin{equation}
  \sqrt{\langle\psi|\hat{h}^2|\psi\rangle - \langle\psi|\hat{h}|\psi\rangle^2}.
\end{equation}
The resulting wave functions are stored in the output file given by {\tt wffile} in the above input file, i.e. {\tt O18} in this case.

\section{Dynamic calculations and code extension}
The version of \textit{Sky3D} that comes from the Computer Physics Communications library includes the ability to apply an external field, but the only multipole operator is the quadrupole $Q_{20}$ tensor.  Here, an example is given of how to extend the code to include monopole resonances, along with a sample run.

The file {\tt external.f90} contains the relevant code which gives the multipolar boosts to the nucleus.  To adapt it to include monopole resonances, the following steps are needed

First, a new variable should be declared to denote the strength of the monopole kick.  In analogy to the existing strength variable for the $Q_{20}$ kick, {\tt amplq0}, we declare and initialise {\tt amplm=0.0\_db} in the same declaration statement in which {\tt amplq0} is declared.

Second, {\tt amplm} should be added to the {\tt extern} namelist so that the user can input the strength of the monopole boost in the input file.

Third, the boost operator itself should be applied.  The quadrupole boost is set up in the initialisation subroutite {\tt getin\_external}, via the generation of the array {\tt extfield} which contains the external field to be applied to the nucleus.  The monopole boost can simply be added to this field, so that in principle one may simultaneously apply both monopole and quadrupole boosts.  The spatial form of the monopole boost is given conventionally as
\begin{equation}
  f_M(r) \propto r^2.
\end{equation}
In the code, the variable {\tt facr} works out the relevant boost for a particular coordinate (inside a nested {\tt DO}-loop over all spatial points).  Separate neutron and proton factors are calculated to deal with isovector and isocalar boosts, which we can make use of.

There are two different options for how the field is applied - \textit{periodic} or \textit{damped}, corresponding to one that matches periodic boundary conditions, and one that is masked around the nucleus via a three-dimensional Fermi function.  Only one method is used in a particular run, and for the purposes of this extension, the monopole is implemented for the damped method.  This is accomplished by changing lines 63 and 64 from

\footnotesize
\begin{verbatim}
facr=amplq0 *(2.D0*z(iz)**2-x(ix)**2-y(iy)**2) &
   /(1.0D0+EXP((SQRT(x(ix)**2+y(iy)**2+z(iz)**2)-radext)/widext))
\end{verbatim}
\normalsize
to
\footnotesize
\begin{verbatim}
  facr=(amplq0 *(2.D0*z(iz)**2-x(ix)**2-y(iy)**2) &
     +amplm * (x(ix)**2+y(iy)**2+z(iz)**2) ) &   
   /(1.0D0+EXP((SQRT(x(ix)**2+y(iy)**2+z(iz)**2)-radext)/widext))
\end{verbatim}
\normalsize

These changes suffice to initialise the nucleus in a monopole kick.  The extent and form of the masking function are determined by the {\tt radext} and {\tt widext} variables. A suitable input file to calculate a isoscalar giant monopole interaction is
\small
\begin{verbatim}
 &files wffile='restart'/
 &force name='SkI4', pairing='NONE' /
 &main mprint=20,mplot=100,
  mrest=100,writeselect='rc',
  imode=2,tfft=T,nof=1 /
 &grid nx=24,ny=24,nz=24,dx=1,dy=1,dz=1,
	periodic=F /
 &dynamic nt=5000, dt=0.2, mxpact=6, texternal=T/
 &extern ipulse=0,isoext=0,amplm=0.001,radext=4.0,
         widext=1.0, textfield_periodic=F /
 &fragments filename=1*'../static/O18',fix_boost=T,
      fcent(1,1)=0,0,0    /
\end{verbatim}
\normalsize

In this input file, we are running in {\tt imode=2}, meaning a time-dependent calculation.  The {\tt \&dynamic} namelist controls the total number of time steps ({\tt 5000}) along with the time step-size ({\tt 0.2 fm/c})

The output of the code most useful for analysing the response of the nucleus to the isoscalar monopole boost are the files containing the various moments, by default called {\tt monopoles.res}, {\tt dipoles.res}, and {\tt quadrupoles.res}.  For the basic analysis, that would e.g. lead to the strength function, one typically follows the same kind of moment as provided the kick.  The Fourier transform of this moment is essentially the strength function for the giant resonance \cite{Alm05b}, as given by linear response theory; 
\begin{equation}
  S(E) = \sum_\nu |\langle\nu|F|0\rangle|^2\delta(E-E_\nu).
\end{equation}

\begin{figure}[tb]
  \includegraphics[width=\textwidth]{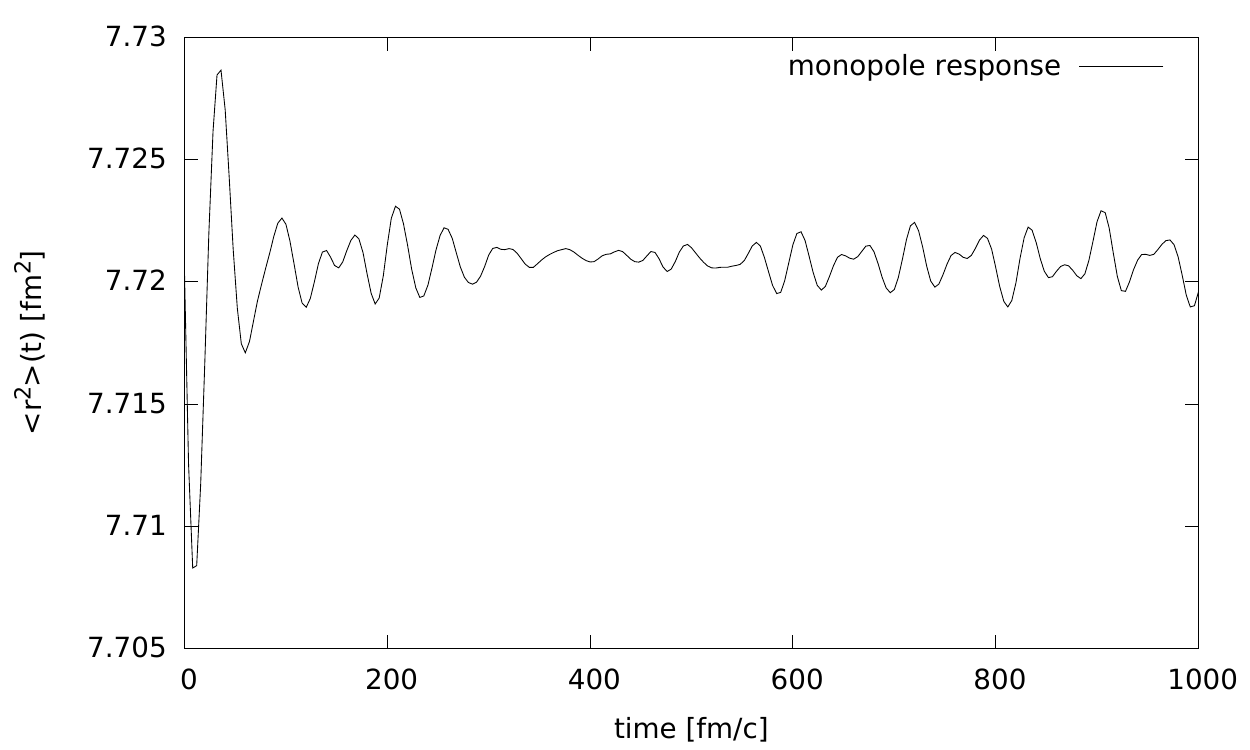}
  \caption{The time-dependent expectiation value of $r^2$ in response to a monopole kick in $^{18}$O.\label{fig:gmrtime}}
\end{figure}

The contents of {\tt monopoles.res} file includes, out of the box, the time, the r.m.s. neutron radius, the r.m.s. proton radius, the matter radius and the isovector radius (i.e. the difference between proton and neutron radii).  The matter radius is what is needed for the linear response, but the output in {\tt monopoles.res} is too low in precision for the purposes of dealing with linear response in the small-amplitude limit.  Of course, one can straightforwardly alter the code to produce enough precision in this file, but in fact the necessary data is available in the {\tt quadrupoles.res} file, whose columns include the expectation values of $x^2$, $y^2$ and $z^2$ separately for protons and neutrons.  For example, the column labelled {\tt x$^2$(n)} gives the quantity $\frac{1}{N}\int d^3r \rho_n x^2$.  If we denote this column $x_n^2$, and similarly for the other columns, then the response to the monopole kick can be calculated as
\begin{equation}
  R(t) = (Nx_n^2 + Ny_n^2 + Nz_n^2 + Zx_p^2+Zy_p^2+Zz_p^2)/A. \label{eq:rt}
\end{equation}
This can hence be conveniently post-processed from the output without modification, and the result of this quantity is shown in figure \ref{fig:gmrtime}.

The figure shows the characteristic short-time damping of a resonant state followed by revenant vibrations after around $t=400$fm/c due to rebounding of outgoing flux from the walls of the coordinate space box \cite{Par14}.

\begin{figure}[tb]
  \includegraphics[width=\textwidth]{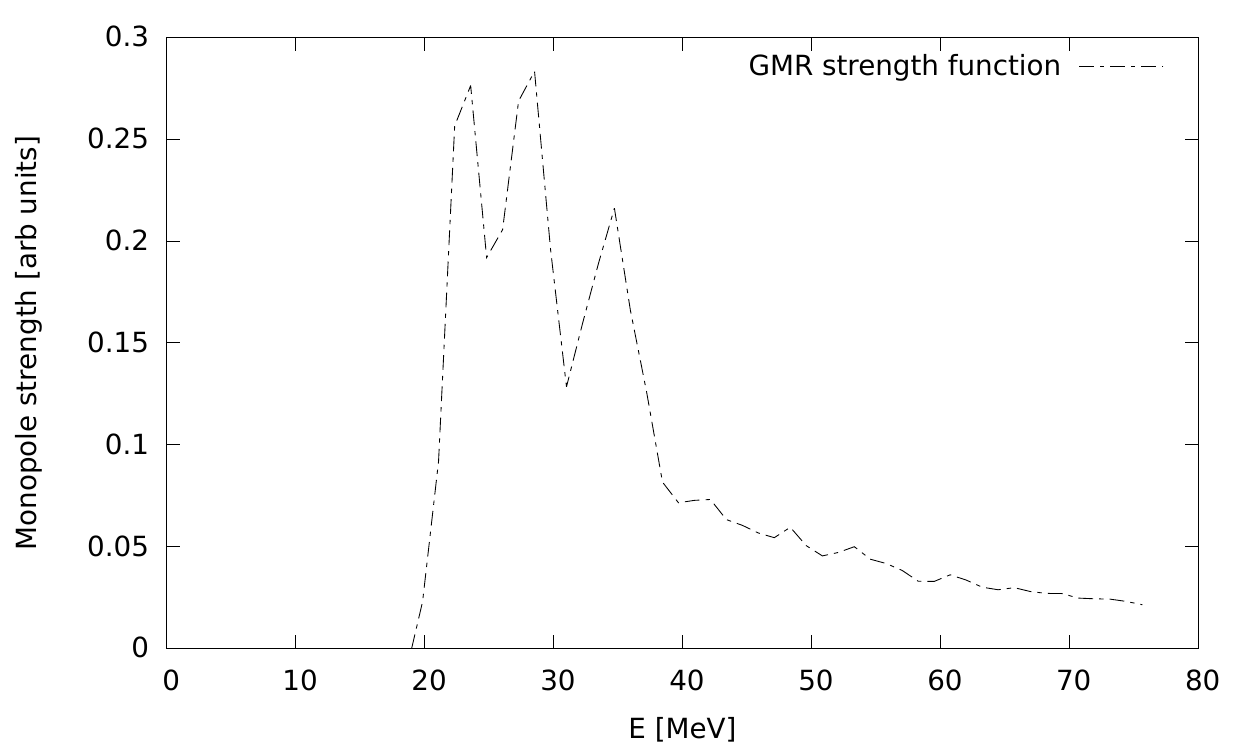}
  \caption{Monopole strength in $^{18}$O using the SkI4 force\label{fig:strength}}
\end{figure}

Finally, one would like to be able to turn the time-dependent response into a strength function.  The \textit{Sky3D} distribution comes with a utiliy program (in the testing foloder) called {\tt spectral\_analysis.f90}.  It turns the time-dependent moment and transforms it to an energy spectrum.  One needs to subtract the $t=0$ value of the moment in order to centre it around zero.  The resulting output of the code gives the strength function, as well as the power spectrum.  A straightforward running of the expression (\ref{eq:rt}) through the {\tt spectral\_analysis} code is shown in figure \ref{fig:strength}

\section{Conclusion}
This paper has reported on some aspects of the \textit{Sky3D} code, in particular giving a practical example of how it can be extended to give useful functionality beyond what is present in the standard distribution, as well as giving an example of post-processing the output of the code.  Extending to other forms of multipole excitations could be performed along very similar lines.  The user manual \cite{Mar14} gives suggestions for other ways to extend the code.
\section*{Acknowledgements}
This work was supported by the UK STFC through grant numbers ST/J000051/1 and ST/L005816/1 and running time granted by STFC on the DiRAC computer cluster.  I gtatefully acknowledge the other members of the Sky3D collaborations; Joachim Marhun, Paul-Gerhard Reinhard and Sait Umar, and also the students and postdocs who have worked with me over the years of developing, testing and running this and related codes: Malcolm Brine, James Broomfield, Emma Suckling, Chris Pardi, Phil Goddard, Daniel Almehed, Sara Fracasso, Ed Simmons and James Petts.

\end{document}